\documentclass[aps,prb,reprint,amsbsy,amsfont,amsmath,floatfix,floats,
   showpacs,showkeys,longbibliography]{revtex4-1}

\usepackage{dcolumn}
\usepackage{bm}
\usepackage{graphicx}
\usepackage{color}

\newcommand*\Rt[1]{$\sqrt{#1}$}

\begin{document}

\title{Incipient triple point for adsorbed xenon monolayers:\\
                        Pt(111) versus graphite substrates}
\author{Anthony D. Novaco}\email[E-mail: ]{novacoad@lafayette.edu}
\affiliation{Department of Physics, Lafayette College \\
                                    Easton, Pennsylvania 18042, USA}

\author{L. W. Bruch}\email[E-mail: ]{lwbruch@wisc.edu}
\affiliation{Department of Physics, University of Wisconsin--Madison\\
                                   Madison, Wisconsin 53706, USA}

\author{Jessica Bavaresco}\email[E-mail: ]{jbavaresco@ufmg.br}
\affiliation{Departamento de F\'\i sica, Universidade Federal de Minas Gerais\\
    Caixa Postal 702, Belo Horizonte, Minas Gerais 30123-970, Brazil}

\date{April 27, 2015}

\begin{abstract}
Simulation evidence of an incipient triple point is reported for xenon
submonolayers adsorbed on the (111) surface of platinum.  This is in stark
contrast to the ``normal'' triple point found in simulations and experiments
for xenon on the basal plane surface of graphite.
The motions of the atoms in the surface plane are treated with standard 2D
``NVE'' molecular dynamics simulations using modern interactions.
The simulation evidence strongly suggests an incipient triple point in the
$120$--$150$~K range for adsorption on the Pt~(111) surface while the
adsorption on graphite shows a normal triple point at about $100$~K.
\end{abstract}



\maketitle

The adsorption of xenon on the (111) surface of platinum (Xe/Pt)
is one of the more interesting cases of physical adsorption.
The potential energy surface is very strongly corrugated and a wide variety
of monolayer structures occur.\cite{BruDieVen07}
The binding to the surface is relatively large, so that the vapor
pressure coexisting with the monolayer remains small enough
at temperatures approaching monolayer melting that experimental probes
such as low energy electron diffraction and helium atom scattering remain
viable. Thus, direct measurements of monolayer dynamics near melting, which
are very scarce in the family of nominally two-dimensional systems, may be
feasible in this case.

The corrugation for Xe/Pt is much larger than it is for xenon on the basal
plane of graphite (Xe/Gr).\cite{BarRet92,*BarRet94,BruNov00,BruNov08,NovBru14}
The Xe/Pt minimum barrier to translation from one adsorption site to the
next is roughly $275$~K, whereas the minimum in the effective
interaction between two Xe atoms is about $238$~K
(for Xe/Gr, the barrier is about $50$~K.).
This should produce strong competition between atom-atom forces and
atom-substrate forces.
However, there is a report\cite{PoeVerCom85} that the Xe/Pt triple
point temperature is $98\pm2$~K, essentially equal to that of
Xe/Gr,\cite{BruDieVen07,NovBru14} a surface with a smaller
corrugation\cite{BruNov00,BruNov08} by a factor of about $5$ or $6$.
While it is expected that thermal excitations will somewhat smooth
the effects of substrate corrugation,\cite{BruNov00} and this
indeed is the case with Xe/Gr,\cite{BruNov08,NovBru14} results of
preliminary molecular-dynamics calculations\cite{Note-Stud} for Xe/Pt
indicate there is insufficient smoothing
to explain such similar triple point temperatures. The
results of those preliminary calculations suggest
there should be a measurable effect upon the triple point of Xe/Pt
as a result of this corrugation.
Dilemma: Why should a system with such strong corrugation behave
so like a system with rather weak corrugation?\cite{NovBru14}

The strong corrugation of the Xe/Pt system and the fact that the monolayer
apparently melts from the commensurate solid phase\cite{PoeVerCom85} suggests
an interesting possibility: the existence of an incipient triple point.
This can occur
if the substrate corrugation is strong enough to sufficiently lower the
free energy of the solid phase so as to maintain a direct
transition from the solid phase into the gas phase, bypassing the
liquid phase altogether.\cite{Nis86,NisGri85}
With Xe/Pt, we not only have strong corrugation,
but also a dilated Xe lattice which weakens the
effects of the attractive region of the Xe-Xe interaction.  This dilation
is due to the $\sqrt{3}\!\times\!\sqrt{3}$-R$30^{\circ}$ phase
(\Rt{3} phase) having a lattice spacing\cite{KerDavZep88} of
$4.80$~\AA, which is significantly
larger than the ``natural'' lattice spacing of the Xe,
that being in the $4.38$--$4.55$~\AA\ range.\cite{NovBru14} 
This implies Xe atoms, when they are located on
the \Rt{3} adsorption sites, tend to be kept away from the hard-core
region of their mutual interaction and found in the weaker region
of their attractive well.  Therefore, there exists the possibility that
this system will, at high temperatures, act much like a lattice gas on
a triangular lattice (here, exhibiting repulsive nearest neighbors and
attractive next-nearest neighbor interactions). This is an interesting
possibility since transitions in similar systems have been examined
theoretically,\cite{SchSid73} including renormalization group
calculations.\cite{SchWalWor77,SchWalWor77a}
The results reported in this communication are part of a much larger
study\cite{Note-NovBav} of the structure and thermodynamics of Xe/Pt.
The results reported here for Xe/Gr build upon the analysis
reported in Ref.~{\onlinecite{NovBru14}}.

The simulation model for Xe/Pt is the Xe/Gr model\cite{NovBru14}
adapted to Xe/Pt with new parameters for the substrate-mediated
interaction and a much larger potential
energy corrugation.\cite{BruNov00,BruGraToe00}
Validation of the resulting model for Xe/Pt is mainly based on
the analysis\cite{BruGraToe00} of the stability and dynamics of a
compressed triangular lattice of Xe/Pt with nearest neighbor
spacing $4.33$~\AA.
The Barker-Rettner (BR) model\cite{BarRet92,*BarRet94} is used for the
3D potential energy surface of Xe/Pt as in Ref.~\onlinecite{BruNov00}.
In this work, the BR interaction is simplified by first using a 2D version
of the 3D Steele expansion,\cite{Ste74} and then truncating the
sum in reciprocal space (consistent with the substrate's $6$-fold symmetry).
Thus
\begin{equation}
\label{SteeleEq}
U(\bm{r}) = U_{(1,0)} \sum_{\bm{G}} \exp ( i \bm{G} \cdot \bm{r} ) ,
\end{equation}
where $U$ is the potential energy of the Xe atom in the field of the
substrate, $\bm{r}$ is the displacement vector of the atom
parallel to the surface, and $\bm{G}$ is a member of the set of the six
shortest reciprocal lattice vectors defined for the surface lattice.

The simulations carried out are standard NVE molecular dynamics simulations
in 2D (here NVE being the 2D version: Number-Area-Energy).
The implementation of the basic simulation is exactly
as per Refs.~\onlinecite{BruNov08} and \onlinecite{NovBru14}, using
essentially the same set of parameters.  As in those works, we examined both
constrained geometries (a single phase filling the simulations box) and
unconstrained geometries (an isolated patch surrounded by vapor).
The density of the \Rt{3} structure is denoted
by $\rho_{p}$ for Xe/Pt and by $\rho_{g}$ for Xe/Gr. Simulations for the
unconstrained geometries are for average densities $\rho$ that are roughly
half that of the corresponding \Rt{3} phase.
In addition to the thermodynamic quantities calculated in
Ref.~\onlinecite{NovBru14} (such as the hexatic order parameter $\psi_6$),
calculations are carried out for the specific heat at constant area
$c_a$ and a
second order parameter: the ``Net-Domain-Phase'' (NDP) order parameter.
The set of case studies reported here are detailed in
Table~\ref{Table:CaseStudies}.

Calculations for $c_a$ use the
fluctuations in the kinetic energy of the system.\cite{LebPerVer67}
As a consequence of this approach, there are some problems in evaluating
$c_a$ when the drift in the energy is too large
over the averaging time interval, as this can result in negative values
for $c_a$. Nevertheless, most of the calculations result in values
that are consistent with the slope of the total energy versus temperature
data.\cite{Note-Deriv} However, the latter approach also generates some
values which are negative (resulting from statistical uncertainty in the
thermodynamic quantities).  Negative values for $c_a$ (as well as
some very large positive values) are not included in the plots.

The NDP order parameter is important in determining and understanding the
nature of the order-disorder transition in Xe/Pt.  It is defined by:
\begin{equation}
\psi_0 = \frac{1}{N} \sum_{i=1}^N \exp ( i \bm{\tau}_{(1,0)} \cdot \bm{r}_i ) ,
\end{equation}
where $N$ is the number of Xe atoms and $\bm{\tau}_{(1,0)}$ is a primitive
reciprocal lattice vector for the Xe \Rt{3} structure.
This order parameter is, of course, nothing more than
the basic structure factor evaluated at $\bm{\tau}_{(1,0)}$
and is a sensitive test of the lattice gas ordering in the \Rt{3}
commensurate lattice.  If all the atoms are placed on the ideal lattice
sites of a \Rt{3} phase single domain, this order parameter takes on a
value in the set: ($1.0$, $e^{i 2 \pi /3}$, and $e^{- i 2 \pi /3}$);
the particular value
depends upon which of the three possible site types (sub-lattices) is
occupied.  The NDP order parameter is zero if each adsorption site type is
populated with equal probability, even if that phase is not one of true
disorder, e.g.\ the hexagonal incommensurate phase (HIC) or the striped
incommensurate phase (SI).

\begin{table}
\caption{\label{Table:CaseStudies} Parameters that define the various
case studies discussed in this paper.  All energy values are in kelvin.}
\begin{ruledtabular}
\begin{tabular}{ l  c  c  c  d }
Case Study & Substrate & Projection & Size & U_{(1,0)} \\
\hline
U36-64K\footnotemark[1] & Pt~(111) & U36 & 64K &  -35.6 \\
U36H-20K\footnotemark[2] & Pt~(111) & U36H & 20K &  -35.6 \\
U15H-20K\footnotemark[2] & Pt~(111) & U15H & 20K &  -15.0 \\
U6H-78K\footnotemark[3] & Graphite & U6H & 78K &  -6.0 \\
\end{tabular}
\end{ruledtabular}
\footnotetext[1]{Constrained geometry with 65536 adatoms.}
\footnotetext[2]{Unconstrained geometry with 20064 adatoms.}
\footnotetext[3]{Unconstrained geometry with 78000 adatoms.}
\end{table}

Simulations of constrained geometries are used to determine the
stable low temperature phase for the classical system and in the
interpretation of the high temperature behavior for a parallel set of
unconstrained geometries.  The constrained geometries are also used to test
the sensitivity of the simulations to system size and to follow the
system in a simpler context (having only one phase present at any given
temperature).  However, the transition
temperature for the constrained geometries is quite high, and
not relevant to the experimental conditions. In fact, it would be expected
that layer promotion would become quite important well before the transition
would be reached for these constrained geometries.
Simulations of the unconstrained geometries are used to examine
both the thermal behavior and the structural properties of the submonolayer
patch. These configurations  generally show 2-phase coexistence of a 2D gas and
a 2D dense phase (most cases follow the 2D sublimation curve).

The 2D projection (at very low temperatures) of the BR model for
Xe/Pt gives\cite{Note-U_G}
$ -36$~K $\le U_{(1,0)} \le -34$~K.  This is obtained by assigning
the BR energy barrier to the corrugation given by
Eq.~(\ref{SteeleEq}).  For Xe/Pt, this corrugation produces a \Rt{3}
structure for the ground state.  However, simulations carried out for
smaller corrugations test both the stability of the \Rt{3} ground state
to variations in the corrugation and address the issue of the effective
corrugation being dependent on temperature due to the thermal motion of the Xe
in the direction normal to the surface.\cite{SeyCarDie99,BruNov00,Note-NovBav}
Thermal vibration of the adatom normal to the surface causes the effective
corrugation to decrease as the temperatures
increases.\cite{BruNov00,Note-NovBav}
Although we have examined a number of corrugation values,
we report only those listed in
Table~\ref{Table:CaseStudies}.  While the HIC structure is the ground state
for the smaller corrugation, both high and low corrugation cases for the
unconstrained geometries exhibit a stable \Rt{3}
structure below melting. This behavior is independent of the
different initialization structures (\Rt{3} or HIC)
generated for the system. However, for the U15H case the transition from
HIC to \Rt{3} occurs just below melting.

\begin{figure}
   \centering
   \includegraphics[width=2.8in]{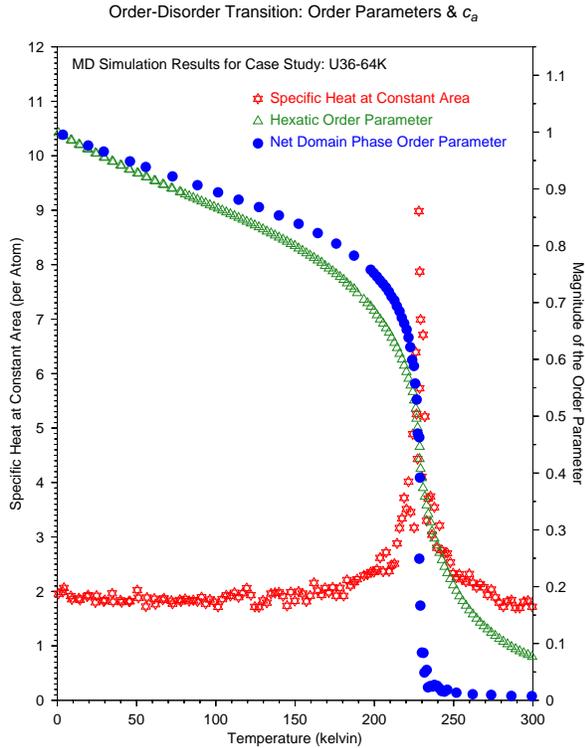}
   \caption{\label{OrderDisorderPlots-1}(Color online) The hexatic
    order parameter, the NDP order parameter, and the specific heat
    for case study U36-64K of Xe/Pt, showing the alignment of
    the peak in the specific heat with the sharp drop in value of $\psi_0$.
    The density is $\rho_p$.}
\end{figure}

Examination of the heating data for case study U36-64K shows a distinct
transition from an ordered to a disordered state at about $230$~K.
Figure~\ref{OrderDisorderPlots-1} shows the specific heat and the magnitudes
of the order parameters $\psi_0$ and $\psi_6$.  The specific heat shows
clear signs of a transition, with a moderate to large increase in the its
value and a classic $\lambda$ shape.  As the temperature is increased,
$\psi_0$ clearly shows a sharp drop in its value while
$\psi_6$ displays a more gradual decrease in the orientational order of
the system.  The specific heat peak is aligned very nicely with the sharp
decrease in $\psi_0$ and suggests that $\psi_0$ is a good
measure of this order-disorder transition.  Furthermore, the specific heat
is more suggestive of a classic continuous transition than of the
traditional discontinuous transition of a triple point constrained by a fixed
area (which would have a trapezoidal profile).  It is relevant here
to be mindful that the natural spacing of the Xe atoms is smaller than the
spacing in the \Rt{3} structure. Thus the usual situation of a triple point
being associated with an increase in area is not relevant and the
structure seems to be dominated by the corrugation of the substrate even at
the (incipient) triple point.

\begin{figure}
   \centering
   \includegraphics[width=2.8in]{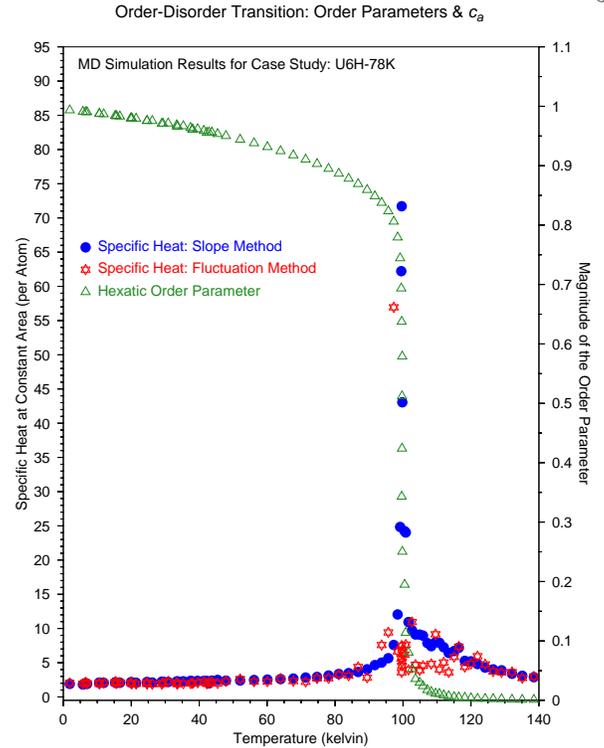}
   \caption{\label{SpHeat.vs.Temp-Xe.Gr}(Color online) The hexatic
    order parameter and the specific heat
    for case study U6H-78K of Xe/Gr, comparing methods for the
    calculation of the specific heat.
    The average density is about $\rho_g/2$.}
\end{figure}

The results for Xe/Pt are different from those obtained for Xe/Gr, where
there is clear evidence, within statistical uncertainty, of a vertical rise
in the energy at a fixed temperature.\cite{NovBru14} This results in
a specific heat that has a sharp (almost vertical) rise in the specific
heat and a much smaller precursor to this sharp rise as seen
in Fig.~\ref{SpHeat.vs.Temp-Xe.Gr}, where the hexatic order parameter
shows a sharp drop in its value. This drop is well aligned with
the vertical rise in the total energy, implying that $\psi_6$ is a
good measure of the order-disorder in the Xe/Gr system.\cite{NovBru14}

\begin{figure}
   \centering
   \includegraphics[width=2.8in]{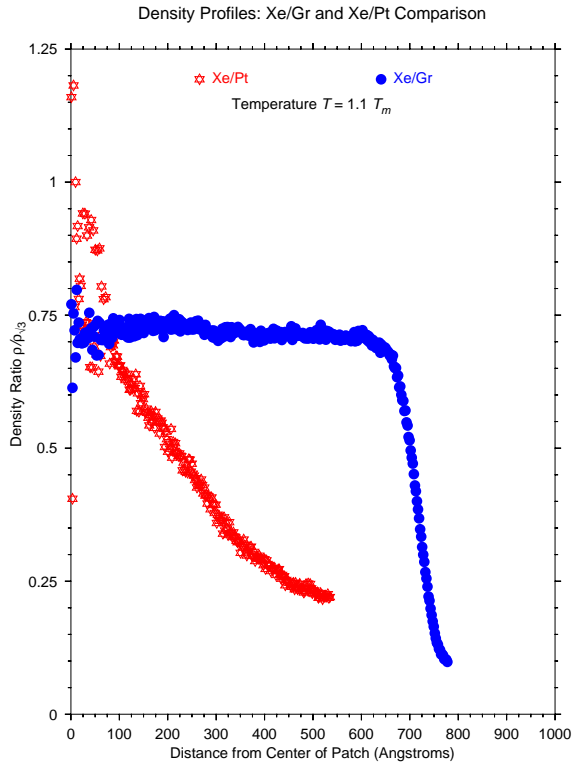}
   \caption{\label{DensityProfiles}(Color online) Comparison of
    the density profiles for patches of Xe/Gr
    (case study U6H-78K; $T_m \simeq 100$~K;
         $ \rho \simeq \frac{1}{2} \rho_g $) and Xe/Pt
    (case study U36H-20K; $T_m \simeq 150$~K;
    $ \rho \simeq \frac{1}{2} \rho_p $)
    just above the melting (order-disorder) transition.}
\end{figure}

All these behaviors are consistent with structure factor plots of
each system showing a clear loss of order as the system moves through the
corresponding transition.\cite{NovBru14,Note-NovBav}
However, for Xe/Pt, there is no clear indication
of a self-bound liquid state in the spatial plots of the system as there
is\cite{BruNov08,NovBru14} for Xe/Gr.
Figure~\ref{DensityProfiles} shows a distinct liquid-gas interface for Xe/Gr
just above its triple point, but no hint of such an
interface for Xe/Pt just above its order-disorder
transition. In both cases, these plots are for initially
hexagonal patches centered in the simulation cell.
For Xe/Pt, there is no evidence of a discontinuous transition; rather there
is evidence of a likely continuous one. For the unconstrained geometries,
there tends to be more scatter in the data, and both order parameters drop
more gradually to zero (as compared to the constrained geometries).
For U36H-20K, the transition temperature is $\simeq 150$~K,
significantly lower than that for U36-64K, but still significantly
higher than the experimental value\cite{PoeVerCom85}
of $\simeq 100$~K.

\begin{figure}
   \centering
   \includegraphics[width=2.8in]{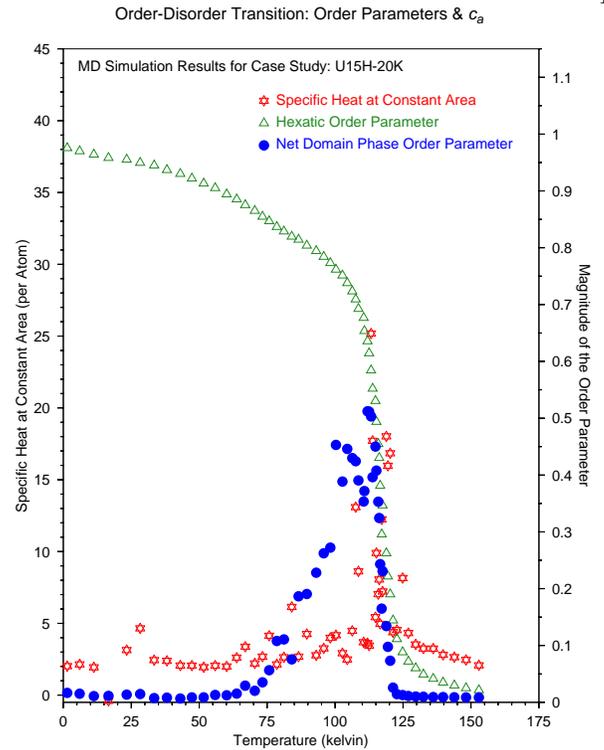}
   \caption{\label{OrderDisorderPlots-2}(Color online) The hexatic
    order parameter, the NDP order parameter, and the specific heat
    for case study U15H-20K of Xe/Pt, showing the
    alignment of the peak in the specific heat with the
    sharp drop in value of $\psi_0$.
    The average density is about $\rho_p/2$.}
\end{figure}

At temperatures near the transition temperature, the corrugation
is smoothed by the thermal motion perpendicular to the surface (lateral
motion smoothing is part of the MD simulation). Preliminary
estimates of smoothing\cite{Note-NovBav} at these high temperatures
result in $|U_{(1,0)}|$
values in the $20$--$25$~K range, corrugations which still
produce order-disorder
temperatures higher than the experimental value. Question: How large
must the smoothing be to bring the simulation results closer to the
experimental temperature?  In Fig.~\ref{OrderDisorderPlots-2}, the effects of
extreme smoothing are shown using a $U_{(1,0)}$ of $-15$~K. The transition
temperature is about $120$~K, significantly lower than that of the
U36H-20K case study.

The $\psi_0$ order parameter appears to be a better measure than
$\psi_6$ for the order-disorder
associated with the transition from solid to fluid in the Xe/Pt system.
The drop in the order parameter with increasing temperature near the
transition is steeper and more complete for $\psi_0$ than it is for $\psi_6$.
Furthermore, $\psi_0$ is more sensitive than is $\psi_6$ to system
size,\cite{Note-NovBav}
and this is what would be expected if $\psi_0$ is the more thermodynamically
relevant order parameter.  This is in stark contrast to the results for
Xe/Gr,\cite{NovBru14} where $\psi_6$ is a very good measure of
the order-disorder near the triple point of that system as shown in
Fig.~\ref{SpHeat.vs.Temp-Xe.Gr}.

The melting of Xe/Pt is clearly of a different nature than
the true triple point melting of Xe/Gr.  This can be seen by the differences
in the specific heats, order parameters, and structural orderings
of the two systems. These calculations and other observations of the
solid\cite{SeyCarDie99,Thesis-Ward} above $100$~K make it
likely the phenomenon at $98$~K in Ref.~\onlinecite{PoeVerCom85} is
not triple point melting.  These simulations
suggest that a Xe/Pt triple point at $\simeq 100$~K is inconsistent with
the observation of a \Rt{3} phase at $60$~K.
The case of Xe/Pt has all the earmarks of an incipient
triple point as described in Ref.~\onlinecite{Nis86}, having a specific
heat that looks much like that of a lattice gas transition as described in
Ref.~\onlinecite{SchWalWor77a}. Since there are finite-size effects (as there
are in any simulation), it is not possible to rule out a normal triple point
and a normal critical point separated by a small temperature gap.  However,
such effects would also bedevil the experimental systems. Studies on larger
systems having longer run times would be welcome, as well as a better
understanding of the thermal smoothing of the corrugation. More thorough
experimental studies of the dense monolayer at $100$--$125$~K (such as
measurements of diffusive motions) might be decisive in establishing the
way in which the monolayer disorders.
\begin{acknowledgments}
We would like to acknowledge and thank C. Chen and S. Kapita for
the work they did on the preliminary studies that preceded this work.
We also thank Lafayette College for its generous support and the
Computer Science Department of Lafayette College for use of their
research computer cluster. Jessica Bavaresco's exchange visit to Lafayette
College during the $2012$ calendar year was sponsored by the Brazilian
government agency CAPES as part of the Science Without Borders program.
\end{acknowledgments}


\begin{thebibliography}{22}
\bibitem[{Not({\natexlab{c}})}]{Note-Deriv}
\bibinfo{note}{The slope is determined by a numerical derivative of the total
  energy, using second order fits to sequencial sets of data triplets.}
\bibitem[{Not({\natexlab{d}})}]{Note-U_G}
\bibinfo{note}{The exact value depends upon the calculational details. These
  include: how to assign the corrugation to a single Fourier amplitude, what
  approximations are made for the variation of the Fourier amplitudes in the
  direction normal to the surface, and how the zero-point energy associated
  with that direction is treated.}
\end{thebibliography}
\end{document}